
\input harvmac.tex

\def\frac#1#2{{#1\over#2}}

\def\journal#1&#2(#3){\unskip, #1~\bf #2 \rm(19#3) }
\def\andjournal#1&#2(#3){\sl #1~\bf #2 \rm (19#3) }

\def\det{{\rm det}}
\def\exp{{\rm exp}}

\catcode`\@=11
\def\slash#1{\mathord{\mathpalette\c@ncel{#1}}}
\overfullrule=0pt
\def\steepslash{\c@ncel}
\def\frac#1#2{{#1\over #2}}

\def\inbar{\,\vrule height1.5ex width.4pt depth0pt}
\def\IB{\relax{\rm I\kern-.18em B}}
\def\IC{\relax\hbox{$\inbar\kern-.3em{\rm C}$}}
\def\IP{\relax{\rm I\kern-.18em P}}
\def\IR{\relax{\rm I\kern-.18em R}}
\def\IZ{\relax\ifmmode\mathchoice
{\hbox{Z\kern-.4em Z}}{\hbox{Z\kern-.4em Z}}
{\lower.9pt\hbox{Z\kern-.4em Z}}
{\lower1.2pt\hbox{Z\kern-.4em Z}}\else{Z\kern-.4em Z}\fi}

\def\ddz{\partial_z}

\catcode`\@=12
\def\tr{{\rm tr}}

\Title{IASSNS-HEP-92/61}{Correlation Functions\ in the Itzykson-Zuber  Model}
\centerline{Samson L. Shatashvili\footnote{$^\dagger$}
{Research  is supported  by DOE grant DE-FG02-90ER40542.}\footnote{$^*$}
{On leave of absence from SPOMI, Fontanka 27, St.Petersburg 191011,
Russia.}}
\bigskip\centerline{School of Natural Science}
\centerline{Institute for Advanced Study}
\centerline{Olden Lane,Princeton,NJ 08540}
\vskip .5in

The n-point function for the integral over unitary matrices with
Itzykson-Zuber measure is reduced to the integral over Gelfand-Tzetlin
table; integrand (for generic n) is given
by linear exponential times rational function. For n=2 and in some cases for
$n>2$
later in fact is the polynomial and this allows to give an explicit and simple
expression for all 2-point
and a set of n-point functions. For the most general n-point
function a simple linear
differential equation is constructed.

\Date{September, 92}

\newsec{Introduction}

In this letter I'll consider the following correlation function :

\eqn\un{<g_{i_1j_1}g^+_{k_1l_1}...g_{i_nj_n}g^+_{k_nl_n}>=
\int\left[d\mu(g)
\right]\exp[\Tr(gMg^+N)]g_{i_1j_1}g^+_{i_1j_1}...g_{i_nj_n}g^+_{k_nl_n}.}
Here g is the N dimensional unitary matrix and M and N are
Hermitian. Measure
of integration is Haar measure.Without lack of generality
we could assume
that $M$ and $N$ are diagonal.

For the case of n=0 (partition function)
this integral was calculated
by Harish-Chandra \ref\hch{Harish-Chandra, Amer.J.Math.,
79(1957),87.
                       }
and Itzykson and Zuber \ref\iz{
K.Itzykson and  J.B.Zuber, J.Math.Phys.,21, (1980), 411.
          }
long time ago. Here we will use the method previously used
in similar problem
in \ref\afs{A.Alekseev,L.Faddeev and S.Shatashvili,Journal
of Geometry and
Physics, (1989)v.3.              } ; this simple algebraic
method
is known in literature on representation theory
since 1950 \ref\gn{I.M.Gelfand and M.A.Naimark, Trudi MIAN,
v36,(1950).}.
Let me mention that the  main motivation to look on
the integral \un\     is related to
investigation of Kazakov-Migdal
model \ref\km{V.Kazakov and A.Migdal, Induced QCD at Large $N$,
Paris-Princeton
preprint LPTENS-92/15/PUPT-1322, May,
1992.}; also , this kind of integrals might be
interesting for string
theory related matrix models \ref\mrd{M.Douglas,
unpublished.}.

I'll show that the integral \un\ can be written in the form:

\eqn\uns{ < g_{i_1j_1}g^+_{k_1l_1}...g_{i_nj_n}g^+_{k_nl_n} >=
\frac{C_N}{\Delta(M)}\int\prod^{N-1}_{k=1}\prod^{N-k}_{i=1}dm^i_k
 R{[m]}\exp[\sum^N_{k=1}(\sum^{N-k+1}_{i=1}m^{k-1}_i-\sum^{N-k}
_{i=1}m_i^k)N_k].}
where $\Delta(M)$ is the Vandermond determinant constructed
from eigenvalues
of matrix $M, m^i_k$ are Gelfand-Tzetlin coordinates
defining the convex
body (see fig.1)

\eqn\gt{ m^i_k > m^{i+1}_k > m^i_{k+1}, }
with $m^0_k$ being the eigenvalues of matrix M (we assume
that they are
ordered:
${m^0_1 > m^0_2 > ... >m^0_N}$ ), $N_k$ are eigenvalues
of matrix $N$, $C$ is the number (see (2.12) )
and $R{[m]}$ is given (in general case ) by
 rational function on GT table. This function will be described
explicitly in section 3
 together with linear differential equation for \uns\
(I'll give the
corresponding algorithm).

$R[m]$ reduces to simple polynomial in the case of
$i_1=i_2=...=i_n=1$ and this
allows to give the explicit formula (we denote by
$\Delta(N_2,...,N-N)$ the
Vandermond determinant for eigenvalues $N_2,N_3,...,N_N$):

\eqn\main{\eqalign{&<g_{1j_1}g^+_{k_11}g_{1j_2}g^+_{k_21}
...g_{1j_n}g^+_{k_n1}>=
\delta_{j_1k_1}\delta_{j_2k_2}...\delta_{j_nk_n}\frac{C_N}
{\Delta(M)\Delta(N_2,. ..,N_N)}\cr
&\int\prod^{N-1}_{k=1}dm^1_k
\frac{\prod^{N-1}_{l=1}(m^1_l-m^0_{j_1})}{\prod_{l\neq{j_1}}
(m^0_l-m^0_{j_1})}\frac{\prod^{N-1}_{l=1}(m^1_l-m^0_{j_2})}
{\prod_{l\neq{j_2}}(m^0_l-m^0_{j_2})}...
\frac{\prod^{N-1}_{l=1}(m^1_l-m^0_{j_n})}{\prod_{l\neq{j_n}}
(m^0_l-m^0_{j_n})}\cr
&\exp[(\sum^N_{k=1}m^0_k-\sum^{N-1}_{k=1}m^1_k)N_1]
\det[exp(m^1_iN_j)]}.}
Here $i=1,2,...,N-1; j=2,3,...,N$.
All nonzero 2-point functions  $< g_{ik} g^+_{ki} >$ and
a set of multi-point
functions $<g_{ij_1}g^+_{k_ni}...g_{ij_n}g^+_{k_ni}>$
are obtained from \main\
by permutation of eigenvalues $m^0_i$ after integration
over $m^1$ in
\main .
\foot{Integral
in \main\ is easy to calculate,but I don't think that
integrated version looks
simpler than this.}

\newsec{Gelfand-Tzetlin parametrization}

Let us denote by X the combination that enters in IZ measure:

$$X=g M  g^+.$$
X is the element of coadjoint orbit of unitary group;
this orbit is labeled by
eigenvalues of M : $m^0_1 > ...> m^o_N $ . We will now
introduce the
coordinates on group and on G/H ,with H being the Cartan
subgroup defined by
M ,
so called Gelfand-Tzetlin coordinates (for the reason why we
call this coordinates GTC see \afs ).

Let $a_i$ be the basis in N dimensional complex plane,
$C^N$, and $e_i$ be
fixed basis , in which matrix M
is diagonal with ordered eigenvalues ${m^o_i}$. Than we
can write

\eqn\eucl{M = \sum^N_{i=1} m^0_i e_i e^+_i,
X = \sum^N_{i=1} m^0_i a_i
a^+_i,
g_{ji}= (e_i,a_j).}
$a_i=ge_i$, ( , ) is the scalar product:
$({\alpha}x,y)=\alpha^*(x,y);(x,{\alpha}y)={\alpha}(x,y).$
Haar measure could be parametrized by the vectors $a_i$, $d\mu(g)=
d\mu(a_1,...,a_N)$. Then

$$ d\mu(a_1,...,a_N) =
\frac{(N-1)!}{(2\pi)^N}dt_1...dt_{N-1}d\theta_1...d\theta_N
d\mu(a_2,...,a_N), $$

\eqn\dt{t_k= |g_{1k}|^2, k=1,...,N-1;}

$$ \theta_k = \arg g_{1k}, k=1,....,N.$$

One can move to the new variables $m^1_i$ from $t_i$
in following way:
suppose
$ m^1$ is the eigenvalue of matrix $PMP$, where $P$ is
the projector on the subspace, orthogonal to vector $a_1$.
This means that if $f$
is
corresponding eigenvector, $f=\sum^N_{k=1}b_ke_k$,one has

$$ PMf=mf ,    Mf=m^1f+ba_1,   f=\sum^N_{k=1}b_ke_k.$$
Taking the scalar
product of $Mf$ with $f_k$ and $e_k$ we get
$$b_k=\frac{bg_{1k}}{m^0_k-m^1},b^2=\sum^N_{k=1}
\frac{|g_{1k}|^2}{(m_k-m)^2}.$$
{}From the orthogonality condition,$(f,a_1)=0$,  one obtains
equation, which
relates $g_{1k}$ and eigenvalues $m^1_k$ by

$$\sum^N_{i=1}\frac{t_k}{m^0_k-m^1}=0, t_k=|g_{1k}|^2,$$
\eqn\ort{\sum^N_{k=1}t_k=1.}
This equation describes one-to-one correspondence of coordinate
system $t_k$
and

$m^1_k$; from it immediately follows that eigenvalues
obey inequalities

\eqn\gto{m^0_i > m^1_i > m^0_{i+1},}
and this intervals are filled densely.Moreover after some
simple algebra one
finds
 relation

\eqn\good{t_k=\frac{\prod^{N-1}_{i=1}(m^1_i-m^0_k)}
{\prod_{i\neq{k}}(m^0_i-m^0_k)}.}

For to move to coordinates $m^1$ we also need the Jacobian
for integration
measure. From above expression it could be easily obtained
and finally we
have

\eqn\jac{J({t},{m^1})=\frac{\prod_{k<i}(m^1_i-m^1_k)}
{\prod_{l<p}(m^0_l-m^0_p)},}

$$d\mu(a_1,...,a_N)=\frac{(N-1)!}{(2\pi)^N}J(t,m^1)dm^1_1
...dm^1_{N-1}d\theta_1...d\theta_Nd\mu(a_2,...,a_N).$$

In addition one can show also that following formula holds:

\eqn\best{X_{11}=\sum^N_{i=1}g_{1k}m_kg^+_{k1}=\sum^N_{i=1}
t_km^0_k=\sum^N_{i=1} m^0_i-\sum^{N-1}_{i=1}m^1_i.}

This completes the first step in descent procedure,
which embeds the $N-1$
dimensional unitary group in N dimensional one.
After this step
we have $N-1$ dimensional space spanned by orthonormal vectors
$a_2,...,a_N$, diagonal matrix $M^1=PMP$ with eigenvalues
$m_1^1,...,m_{N-1}^1$
and eigenvectors $f$ . It is simple to continue the
descent further;
let me give
the
necessary expressions after the descent procedure is
completed down to the
level
0 \foot{one can derive all this expressions using
the procedure described above, or (after some modifications)
they could be
extracted from \gn, see also \afs } ($f^0_{j_0}=e_k,
\alpha^0_{j_{0}}=1$):

\eqn\imp{\eqalign{g_{lk}=&(e_k,a_l)=\sum^{N-l}_1
\sum^{N-2}_1...\sum^{N-l+1}_1(e_k,f_{j_1}^1)(f^1_{j_1},
f^2_{j_2})...\cr..&(f^{l-1}_{j_{l-1}},a_l)=\sum_{[j_q]}
\prod^{l-1}_{q=0}\frac{\alpha^q_{j_q}(f^q_{j_q},a_{q+1})}
{m^q_{j_q}-m^{q+1}_{j_{q+1}}}},}

\eqn\impo{(a_q,f^{q-1}_k)=\sqrt{t_{kq}}\exp(i\theta_{qk}),}

$$ (f^q_i,f^{q-1}_k)= \alpha^q_i\frac{(a_q,f^{q-1}_k)}
{m^{q-1}_k-m^q_i},$$
where by $\alpha$ and $t$ we have denoted following products

\eqn\im{(\alpha^q_i)^2=-\frac{\prod^{N-q+1}_{j=1}
(m^q_i-m^{q-1}_j)}{\prod^{N-q}_{j\neq{i}}(m^q_i-m^q_j)},}

$$t^q_i=\frac{\prod^{N-q}_{j=1}(m^q_j-m^{q-1}_i)}
{\prod^{N-q+1}_{j\neq{k}}(m^{q-1}_j-m^{q-1}_i)}.$$
The variables $m^k_i$ are (as it follows from descent
procedure, see \gto  )
from Gelfand-Tzetlin table
\eqn\gtm{m^k_i>m^{k+1}_i>m^k_{i+1}}
and $\theta$ are just angle variables,
$\theta^k_i=[0,2\pi]$.
Using all this one could check by straightforward
calculation that

\eqn\cart{X_{kk}=\sum^N_{l=1}g_{kl}m^0_lg^+_{lk}=
\sum^N_{l=1}m^0_l|g_{kl}|^2=\sum^{N-k+1}_{i=1}m^{k-1}_i-
\sum^{N-k}_{i=1}m^k_i, X_{NN}=m^N,}

$$d\mu(g)=\frac{C_N}{(2\pi)^{\frac{(N+2)(N-1)}{2}}}
\frac{1}{\Delta(M)}\prod^{N-1}_{k=1}\prod^{N-k}_{i=1}
dm^i_k\prod^N_{k=1}
\prod^{N-k+1}_{i=1}d\theta^i_k,$$
$$C_N=(N-1)!(N-2)!...2!1!.$$
The meaning of symbols $f^i_k$ is simple; f.e. $f^l_k$
forms the orthonormal
 basis in the space spanned by eigenvectors of matrix
$M^l= diag(m^l_k)$

\newsec{Correlation Functions}

Having at the hand the expressions  \imp ,\impo ,\im ,\cart\
it is easy to
prove the statements about \un, announced in introduction, i.e.
derive explicit
formulas
for all 2-point functions and for a set with $n>2$ \main\
and present the
integral representation of general n-point function in terms
of GT table \uns.
Also one could give the linear differential equation for
general n-point function . But first I will start with 0-point
function and
will give the simple derivation of IZ formula:
\foot{this is one more derivation of well-known IZ formula \iz ,
but I decide
to include
it here for completeness of  presentation ; also this kind of
derivation  might
be itself useful,i.e. for large $N$ limit.}

\eqn\izo{<1>=\int\left[d\mu(g)\right]\exp[\tr(gMg^+N)];}
integration is over Gelfand-Tzetlin table \gt.
The result follows immediately from the facts we have learned
in previous
section (we need just \cart) and from two simple observations :
i.Integrand
doesn't depends on
angle variables $\theta$, so it is the linear exponential
on GT table:

\eqn\iz{<1>=\frac{C_N}{\Delta(M)}\int\prod_{GT}
\left[dm\right]\exp[\sum^N_{k=1}(\sum^{N-k+1}_{i=1}
m^{k-1}_i-\sum^{N-k}_{i=1}m^k_i)N_k].}
ii.the integration over the bottom coordinate $m^N$of
$$\exp([(m^{N-1}_1+m^{N-1}_2)-m^N]N_{N-1})\exp(m_NN_N)$$ leads to
$$\frac{\det[\exp(m^{N-1}_iN_{N-j+1})]}{N_N-N_{N-1}},$$
and variables $m$
are from the second (counting
from bottom) line;$i,j=1,2$. Now, according \iz , we have to multiply this
expression on exponential
$$\exp(\sum^3_{i=1}m_i^{N-2}-\sum^2_{i=1}m^{N-1}_i)N_{N-2}$$
and integrate over
$m^{N-1}_i$ ; once  again we get the similar result:
$$\frac{\det[\exp(m^{N-2}_iN_{N-j+1})]}
{\Delta(N_N,N_{N-1},N_{N-2})},$$with
$i,j=1,2,3$.
Thus, using the induction method one easily proves that following
formula holds:
\eqn\izf{\eqalign{\int\prod^{l-1}_{i=1}\left[dm^{N-l+1}_i\right]
&\exp{[\sum^l_{i=1}m^{N-l}_i-\sum^{l-1}_{i=1}m^{N-l+1}_i]
N_{N-l+1}}\frac{\det[\exp(m^{N-l+1}_iN_{N-j+1})]}
{\Delta(N_N,...,N_{N-l+2})}=\cr&=\frac{\det[\exp(m^{N-l}_pN_{N-q+1})
 ]}{\Delta(N_N,..,N_{N-l+1})}};}
here $p,q=1,...,l;i,j=1,...,(l-1).$

This completes the derivation of IZ formula in our approach,
based on GT
parametrization:

\eqn\izfi{<1>=\frac{C_N}{\Delta(M)\Delta(N)}\det[exp(M_iN_j)].}
where we have denoted $m^0_i$ by $M_i$.

Let us now move to discussion of correlation functions. First about
general properties:

All nonzero correlation functions in \un\ have $j_m=k_m$
and
$(l_1,...,l_n)=P(i_1,...,i_n)$,where P is the element of
permutation group;
of course the "reversed" statement , when $i_m=l_m$ and
$(k_1,...,k_n)=P'(j_1,...,j_n)$ , is equivalent to above.

To demonstrate this simple fact let us multiply right hand side in \un
by
$$1=\frac{1}{(2\pi)^{2N}}\int\left[dH_1\right]\int\left[dH_2\right],$$
where $H_1,H_2$ are from $U(1)^N$.
Because the IZ measure is invariant under the
transformation $g'=H_1gH_2$ we can perform this
transformation
in the integrand ; thus each operator inside the
correlation function
will be multiplied by $H_1$ from left and $H_2$ from
right (hermitian
conjugates  will be changed correspondingly).
Because the integration
over $H_1,H_2$ is over $U(1)$
angles we see that the integral will be always zero,
except the cases that  stated above.

The expressions derived at the end of section 2
and the statement \izf
immediately leads to prove of \main. We have

\eqn\twopoint{\eqalign{<g_{1j_1}&g^+_{k_11}g_{1j_2}
g^+_{k_21}...g_{1j_n}g^+_{k_n1}>=\frac{C_N}{\Delta(M)}
\int\prod_{GT}dm\frac{\prod^{N-1}_{l=1}(m^1_l-m^0_{j_1})}
{\prod_{l\neq{j_1}}(m^0_l-m^0_{j_1})}\frac{\prod^{N-1}_{l=1}
(m^1_l-m^0_{j_2})}{\prod_{l\neq{j_2}}(m^0_l-m^0_{j_2})}..\cr
&.\frac{\prod^{N-1}_{l=1}(m^1_l-m^0_{j_n})}{\prod_{l\neq{j_n}}
(m^0_l-m^0_{j_n})}\exp[\sum^N_{k=1}(\sum^{N-k+1}_{i=1}
m^{k-1}_i-\sum^{N-k}_{i=1}m_i^k)N_k]=\frac{C_N}
{\Delta(M)\Delta(N_2,...,N_N)}\cr
&\int\prod^{N-1}_{k=1}dm^1_k\frac{\prod^{N-1}_{l=1}(m^1_l-m^0_{j_1})}
{\prod_{l\neq{j_1}}(m^0_l-m^0_{j_1})}
\frac{\prod^{N-1}_{l=1}(m^1_l-m^0_{j_2})}{\prod_{l\neq{j_2}}
(m^0_l-m^0_{j_2})}...\frac{\prod^{N-1}_{l=1}(m^1_l-m^0_{j_n})}
{\prod_{l\neq{j_n}}(m^0_l-m^0_{j_n})}\cr
&\exp[(\sum^N_{k=1}m^0_k-\sum^{N-1}_{k=1}m^1_k)N_1]
{\det[exp(m^1_iN_j)]}\delta_{j_1k_1}...\delta_{j_nk_n}}.}
Here $i=1,2,..,N-1;,j=2,...,N$. As I have mentioned in
introduction the last
integral
is easy to calculate; it is clear form integral
representation that the result
is the sum of quadratic exponentials of eigenvalues
$M_i,N_i$ times rational functions and it is regular
when two eigenvalues of
$M$ coincide.\foot{ we have used the coordinates tied
with matrix $M$ but we
could use in the same integral the GT coordinates related
to matrix $N$; result
shouldn't
depend on this choice; note that GT coordinates for $M$
eigenvalues of $N$
aren't ordered.} So we could permute the eigenvalues of
$M$ to obtain the all
other correlation functions $<|g_{ik}|^2>$, as well as
all n-point functions
with
$i_1=i_2=...=i_n$ (and others that are related to later
by interchanging matrix
$M$ with matrix $N$) after last integration in \twopoint .

Now we will discuss the most general n-point function,
$n>2$, the one that
couldn't be reduced to \twopoint, and will prove \uns.
Using general property discussed above we will order the
correlation function
\un
by the set of indices $(i_1,...,i_n)$,$i_1>i_2>...>i_n$.
According \imp\ we will draw the
two copies of Yung type diagram : one for set
$(i_1,..,i_n;j_1,...,j_n)$
another for $(i_1,...,i_n;k_1,...,k_n)$ (see fig.2).
Each line in the diagram
has length $i_q$, and thus we write numbers
$(i_1,...,i_n)$ in the boxes
at the right end of each line in each diagram.
In the first diagram we put
numbers $(j_1,...,j_n)$ in most left boxes
(again from up to down, but now the
numbers aren`t necessarily
ordered); equivalently in the second diagram we put
at the same place numbers
$(k_1,...,k_n)=P(j_1,...,j_n)$. So each line in first
diagram has $i_q$
in the right and $j_q$ in the left, at the same time the
same line on second
diagram has $i_q$ at the right and $P(j_q)$ at the left.
Now we will fill
all other boxes in the first diagram according following
procedure : pick up
all lines with length $i_n$ (some of the numbers from set
$i$ might be equal,
so we will have several lines with the same length) and
put next from $i_n$
the numbers $(\rho^1_{i_n-1},...,\rho^{A_{i_n}}_{i_n-1})$
from up to down with
requirement that  $$1<\rho_{i_n-1}<N-i_n+1.$$ $A_{i_n}$
is the number of
lines of length $i_n$.We continue this procedure before
reaching the point
with $i_n-p=i_{n-1}$,where p counts the boxes from right
to left on the top
line of diagram. After this the height of each column is
larger,so next
column we will fill with numbers
$(\rho^1_{i_{n-1}-1},...,{\rho_{i_n-1}^{A_{i_{n-1}}}}),$
with $A_{i_{n-1}}$ being the number of lines of length
$i_n$ plus the number
of lines of length $i_{n-1}$.This procedure we will
continue to the end, when
all boxes in Yung diagram are filled. The numbers
$\rho$ should satisfy
unequality
$$1<\rho^A_{q-1}<N-q+1.$$Let us call this diagram
$Y^{ij}(\rho).$ The same
procedure we will repeat
with the second diagram, assigning the numbers
$\rho'$ to empty boxes and defining $Y^{i,P(j)}(\rho')$.

This diagrams are just convenient representation of products

\eqn\one{g_{i_1j_1}...g_{i_nj_n},}
\eqn\two{g^+_{P(j_1)i_1}...g^+_{P(j_n)i_n},}
when they are viewed in terms of sum in \imp .
For to proceed further we
have to assign to each box in the diagram the factor

\eqn\factor{\frac{\alpha^q_{\rho_q}(f^q_{\rho_q},a_{q+1})}
{m^q_{j_q}-m^{q+1}_{j_{q+1}}}}
and then multiply over $q$ in each line; the
same we should do with the second
diagram, but assigning to the boxes the complex
conjugate of \factor.
So, for
each diagram we have assigned the particular product
dictated by above procedure and the diagram. Let us call
these factors $F^{{ij}}(Y)$ and $F^{{iP(j)}}(Y')$.
Finally, for to obtain the factor $R$ coming from
insertion of \one\ and \two\ in
IZ integral
we have to sum over all possible diagrams the objects $F(Y)$
and $F(Y')$ ((with
fixed  $(i_1,...,i_N);(j_1,...,j_N);P$ ) and then multiply:
\eqn\r{R=\sum_{YY'}F(Y)F(Y').}

The described procedure looks very complicated but result
of integration over
angle variables makes it simpler. Let us take the product
of $F_1$ and $F_2$
with
some particular set of $\rho$ and $\rho'$ and integrate
over angle variables
with measure \good . Because we already know that the
dependance on angles
is coming only from pre-exponent in\un (see \good\ and \iz),
this is the only
angle integral we have to calculate. From other
side the angle
dependance in  \factor\ is due to \im, thus we have simple
angle integral
to calculate: fix the column (let say l-th)
in both diagrams and denote by $\Theta$  sum over this
column of the angles $\theta_{\rho_l
,l}$; denote by $\Theta'$ same sum for second diagram.
The angle integral that
enters in \un\ for this column is
just

$$\int\prod_{row[l]}(d\theta)\exp(\Theta- \Theta'),$$
which is zero if the set ${\rho'}$ on the second diagram
in given column
(l-th in this case) is not related to set of ${\rho}$
in this first one by
permutation
$$(\rho'^1_l,...,\rho'^{A_l}_l)=P(\rho^1_l,...,\rho^{A_l}_l).$$

Later relates the numbers in second diagram to numbers
in first by simple law:
we should consider only those pair of diagrams, when
numbers in fixed
column of second are related to numbers in the same
column of first diagram
by some element of permutation group, so for
each diagram Y we have set of
mappings $\tau$ defined by above law: $Y'=\tau(Y)$.
Thus we have for $R$ after
integration over all angle variables:

\eqn\rtwo{R_0=R_0[m]=\sum_{Y,\tau}F^{(i,j)}_0[Y]F^{(i,P[j])}_0[\tau(Y)].}
Here subscript $0$ means that we have to put to zero all angles in
\factor; after this $R$ becomes the
function on GT table, $R_0[m]$
Now it is easy to see that $R_O[m]$ is the rational function.
Simply after
applying the
mapping $\tau$ to $Y$ there are only moduli squares of \impo\
and \im\  left
in\rtwo ; this means that
$R_0[m]$ is rational. On this we complete
the construction of $R$.
Going back to \un in parametrization \cart\ we
have  (for nonzero correlators):

\eqn\fin{\eqalign{<g_{i_1j_1}g^+_{i_1P(j_1)}...
&g_{i_nj_n}g^+_{i_nP(j_n)}>=\frac
 {C_N}{\Delta(M)}\int\prod_{GT}dmR_0[m]
\exp[\sum^N_{k=1}X_{kk}N_k]=\cr
&=\frac{C_N}{\Delta(M)}\int\prod^{N-1}_{k=1}
\prod^{N-k}_{i=1}dm^i_kR_0[m]\exp[\sum^{N-1}_{k=0}
(\sum^{N-k+1}_{i=1}m_i^{k-1}-\sum^{N-k}_{i=1}m^k_i)N_k]}}
and here indices for $R[m]$ are arranged according for those
in right hand
side. Thus we have
derived the formula announced in the introduction.

The integrand in \fin\
still looks complicated, (even having the explicit algorithm
of its
construction),
except the one when $i_1=i_2=..=i_n$; but it seems that it
could be simplified
using algebraic identities for
particular rational functions that enter in $R_0$ \foot{this
should be viewed as conjecture based on knowledge n-point
function \twopoint;
i.e in our basis the expression for $<|g_{ik}|^2>$  has
similar structure as
general correlation function in \fin and at the same time
answer obtained from
the one with $i=1$ by permutation of eigenvalues of $M$ is simple. }.
It would
be also interesting
to have some geometric interpretation of above construction.
At the moment
best we could extract from \fin in general case is following:
first we could
extract from
$R_0[m]$ two polynomials $P_1$ and $P_2$  by defining
\eqn\pp{R_0[m]=\frac{P_1[m]}{P_2[m]}}
\pp uniquely defines $P_1,P_2$; there is a simple formula for
$P_2$ in the case
of  general n-point function $P_2$
\eqn\polin{P_2=\prod_{[q]}\prod_{i\neq{j}}
(m_i^q-m_j^{q-1})^{a_q}\prod_{k\neq{l}}(m_k^q-m_l^q)^{b_q}
\prod_{p\neq{t}}(m_p^{q-1}-m_t^{q-1})^{c_q}}
where a set $[q]$ and numbers $a_q,b_q,c_q$ are fixed
by the particular
correlator
under consideration.
 Then consider more general then \fin  inegral:
introduce for each point on GT table corresponding "dual"
coordinate $z_i^k$
and
define $I$ by
\eqn\gen{I({z})=\int\prod_{GT}dmR[m]\exp[\sum_{k=1}^{N-1}
\sum_{i=1}^{N-k}z_i^km_i^k]}
{}From $I({z})$ integral \fin\  is obtained by requirement
that all coordinate
on the "z" table in given row $k$ are equal to $N_k-N_{k-1}$;
{}From
other side I obeys the linear differential equation
\eqn\differential{P_2({\ddz})I=P_1({\ddz})I_0, I_0=I(R=1),}
so that right hand side in \differential  is known function.
Thus, some particular solution of \differential  for the values
of $z$ , that I
have defined previously,
should coincide with \fin\ and therefore with \un.

\newsec{Conclusion.}

One should note that the expressions \main,\uns\ are
very simple. This
simplicity of course should be related to the fact that
the integral \un\
actually is not over unitary group $G$, but rather over
it's left/right coset.
We have used only part of symmetries when we tied the
coordinate system to
matrix $M$ (and not to matrix $N$).
Using the fact that result shouldn't
depend on this choice additional identities
could be obtained. It might be possible that later
allows to get all nontrivial
n-point functions  for $n>2$ from those that
we have calculated (see \main).

The simplicity of \main, \uns\ might be helpful for
the applications, mainly for
large $N$ limit; later is important both in string
theory and Kazakov-Migdal
model. In this respect I would like
to note the elegancy of \izf; it might lead to
interesting equation in large
$N$ limit.  This results might be  also helpful
in the statistical models
,discussed in\ref\kmsw{I.Kogan,A.Morozov,G.W.Semenoff
and N.Weiss, Area law and
continuum limit in"induced QCD", UBC preprint
UBCTP-92-022, July, 1992.}.
A.Migdal have derived explicit formula for general
2-point function in the
leading order of large $N$
approximation in his  approach based on Riemann-Hilbert
method \ref\m{A.Migdal,
Mixed model of induced QCD,Paris preprint LPTENS-92/23,
August, 1992.}.I know
from A.Morozov \ref\a{A.Morozov, ITEP preprint, to appear.}
that he have
conjectured simple formula for 2-point
function that looks  similar to the one  obtained
from the
\main\ after integration over $m^1$ in the case of
2-point function. It would be
interesting to find the geometric approach that could
explain in more
invariant way the results obtained in IZ model both for
finite and large $N$.

\vskip 1cm

Acknowledgments: I am grateful to
A.Alekseev, E.Brezin, M.Douglas, V.Kazakov, I.Kogan, I.Kostov,
A.Migdal, A.Morozov, A.Niemi, G.Semenoff, M.Staudacher
and N.Weiss for stimulating
discussions. This
research is supported by DOE grant DE-FG-90ER405542.

\listrefs

\end